\documentclass[11pt,twoside]{article}
\usepackage[cp1251]{inputenc}  
\usepackage{graphicx}
\usepackage{amsmath}
\usepackage{amssymb}
\usepackage{mathrsfs}

\begin{document}

\thispagestyle{plain}

\newcommand{\pst}{\hspace*{1.5em}}
\newcommand{\be}{\begin{equation}}
\newcommand{\ee}{\end{equation}}
\newcommand{\ds}{\displaystyle}
\newcommand{\bdm}{\begin{displaymath}}
\newcommand{\edm}{\end{displaymath}}
\newcommand{\bea}{\begin{eqnarray}}
\newcommand{\eea}{\end{eqnarray}}
\newcommand{\bml}{\begin{multline}}
\newcommand{\emltl}{\end{multline}}

\begin{center} {\Large \bf
\begin{tabular}{c}
Evolution Equation for Joint Tomographic Probability \\[-1mm]
Distribution of Spin-1 Particles
\end{tabular}
 } \end{center}
\smallskip
\begin{center} {\bf Ya. A. Korennoy, V. I. Man'ko}\end{center}
\smallskip
\begin{center}
{\it P.N.    Lebedev Physical Institute,                          \\
       Leninskii prospect 53, 119991, Moscow, Russia }
\end{center}
\begin{abstract}\noindent
The nine-component positive vector optical tomographic probability portrait
of quantum state of \mbox{spin-1} particles containing full spatial and 
spin information about the state without redundancy is constructed.
Also the suggested approach is expanded to symplectic tomography representation and 
to representations with quasidistributions.
The evolution equations for constructed vector optical and symplectic tomograms and vector 
quasidistributions for arbitrary Hamiltonian are found.
The evolution equations are also obtained in special case of the quantum system of 
charged \mbox{spin-1} particle in arbitrary electro-magnetic field, which are analogs of 
non-relativistic Proca equation in appropriate representations.
The generalization of proposed approach to the cases of arbitrary spin is discussed.
The possibility of formulation of quantum mechanics of the systems 
with spins in terms of joint probability distributions without the use of
wave functions or density matrixes is explicitly demonstrated.
\end{abstract}


\section{Introduction}
The proposition of optical tomographic description of states of spinless quantum  systems was formulated in
 \cite{BerBer,VogRis}.
Generalising the optical tomography technique the symplectic tomography
was suggested, and evolution equation for symplectic tomograms of
spinless quantum systems was found in \cite{Mancini96,ManciniFoundPhys97}
providing a bridge between classical and quantum worlds (for review see \cite{IbortPhysScr}).  
Evolution equations for optical tomograms of spinless quantum systems
were obtained in Refs.~\cite{Korarticle2,Korarticle1}.

In Ref.~\cite{NewtonYoung} the spin density matrix for particles of arbitrary
intrinsic angular momentum is explicitly expressed in terms of directly 
measurable expectation values of components of multipole moments, 
or the relative weights of partial beams split-up by a Stern-Gerlach
apparatus.

Extending the tomographic approach in Ref.~\cite{DodonovManko1997}
the spin tomography was formulated based on the description
of spin states with the help of positive distribution functions depending 
on continuous variables like Euler's angles, 
a spin state reconstruction procedure similar to the symplectic
tomography was considered, and quantum evolution equation of spin dynamics
was found for continuous spin tomogram.

In Ref.~\cite{Weigert2000} spin dynamics was expressed for 
expectation values of spin projections along a discrete set of fixed directions.
The spin tomography was also studied in 
\cite{Weigert1992,AmietWeigert9,AmietWeigert10,AmietWeigert11,AmietWeigert12,AmietWeigert14,HeissWeigert2000,FilippovManko2010,MankoMarmo2004}
and in other papers.

The first attempt of tomographic formulation of the Pauli equation 
simultaneously describing both spatial and spin dynamics, apparently,
was done in \cite{ManciniOlgaManko2001}. The evolution equation obtained
is extremely complicated, because it uses redundant tomogram 
depending on continuous Euler's angles and on symplectic vareables.

In our resent paper \cite{KorennoyPauli} we introduced the positive vector 
optical tomogram fully describing both spatial and spin characteristics of 
the quantum state of \mbox{spin-1/2} particle
without any redundancy. We obtained the evolution equation for 
this vector optical tomogram and considered  examples of evolution of quantum systems
in proposed representation.  Also we discussed the expansion of our approach to 
representations of Wigner and Husimi quasidistributions (and pointed out the possibility
of dissemination of the discussed scheme  to the Glauber-Sudarshan representation).

The aim of our work is the construction of spin-1 particle quantum state 
vector tomography without redundancy of information 
representing the joint vector distribution for space coordinates and spin projections; and
derivation of the evolution equation for such distribution, which would be an analogue of
non-relativistic Proca equation.  The latter will allow to explicitly demonstrate
the possibility of formulation of quantum mechanics of the systems 
with spins in terms of joint probability distributions without application of
wave functions or density matrixes.

The paper is organized as follows. 
In Sec.~\ref{Sec2} we give basic formulae of tomographic and quasiprobability representations of
quantum mechanics for spinless particles.
In Sec.~\ref{Sec3} we introduce a positive nine-component vector probability and quasiprobability
description of spin-1 particles and give the evolution equations for such  vector-portraits 
of quantum state with arbitrary Hamiltonian.
In Sec.~\ref{Sec4} charged spin-1  in arbitrary electro-magnetic field
is considered in proposed representations, and evolution equations, 
which are analogs of nonrelativistic limit of Proca equation, are obtained.
The conclusion is presented in Sec.~\ref{Sec5}.

\section{Probability representation and evolution \\
of spinless quantum systems} 
\label{Sec2}
Let us review the constructions of  tomographyc or 
quasidistribution representations in general case
for spinless systems. 

If the state of the quantum system is described by the 
density matrix  $\hat\rho$ normalized by the condition $\mathrm{Tr}\hat\rho=1$,
then in accordance with general scheme the tomographic distribution function
or quasidistribution $w(x,\eta)$ is related with the density matrix as follows
(see \cite{Korarticle5}):
\be		\label{equation31_1}
w(x,\eta,t)=\mathrm{Tr}\{\hat \rho(t)\hat U(x,\eta)\},~~~
\hat\rho(t)=\int w(x,\eta,t)\hat D(x,\eta) \mathrm{d}x\,\mathrm{d}\eta,
\ee
where $x$ is a set of distribution (quasidistribution) variables,
$\eta$ is a set of parameters of corresponding tomography, 
and $\hat U(x,\eta)$, $\hat D (x,\eta)$ are dequantizer and quantizer  operators
for appropriate tomographic scheme or corresponding quasidistribution representation.
For Wigner \cite{Wigner32}, Husimi \cite{Husimi40}, and Glauber-Sudarshan \cite{Glauber,Sudarshan}
quasidistributions the set of parameters $\eta$ is absent.

Notion of quantizer and dequantizer is related to star product quantization schemes
(see recent review \cite{SIGMA10(2014)086}).

Quantizer and dequantizer are constrained by the duality relation
\bdm
\mathrm{Tr}\{\hat U(x,\eta)\hat D(x',\eta')\}=\delta(x-x')\delta(\eta-\eta').
\edm
The von-Neumann equation without interaction with the environment 
\be                             \label{vonNeumann}
i\hbar\frac{\partial}{\partial t}\hat\rho=[\hat H,\hat\rho]
\ee
with the help of maps of type (\ref{equation31_1}) transforms to 
evolution equations for tomograms \cite{Korarticle1},
or to Moyal equation \cite{Moyal1949} for the Wigner function \cite{Wigner32}, or to 
evolution equation for other quasidistribution
\be                             		\label{r22_2_22222}
\partial_tw(x,\eta,t)=
\frac{2}{\hbar}\int\mathrm{Im}\left[\mbox{Tr}\left\{\hat H(t)\hat D(x',\eta')
\hat U(x,\eta)\right\}\right]w(x',\eta',t)\mathrm{d}x'\mathrm{d}\eta'\,.
\ee
If we have spinless quantum system in the $N-$dimensional space,
then dequantizer and quantizer for optical tomography
equal
\be		\label{dequantizerOPT}
\hat U_w(\vec X,\vec\theta)=|\vec X,\vec\theta\,\rangle\langle \vec X,\vec\theta\,|=
\prod_{\sigma=1}^N
\delta\left(X_\sigma-\hat q_\sigma\cos\theta_\sigma-\hat p_\sigma\frac{\sin\theta_\sigma}
{m_\sigma\omega_{\sigma}}\right),
\ee
\be		\label{quantizerOPT}
\hat D_w(\vec X,\vec\theta)=\int\prod_{\sigma=1}^N
\frac{\hbar\vert\eta_\sigma\vert}{2\pi m_\sigma\omega_{\sigma}}
\exp\left\{i\eta_\sigma\left(X_\sigma-\hat q_\sigma\cos\theta_\sigma
-\hat p_\sigma\frac{\sin\theta_\sigma}
{m_\sigma\omega_{\sigma}}\right)\right\}
\mbox{d}^N\eta,
\ee
where $m_\sigma$ and $\omega_\sigma$ are constants that have the dimensions
of mass and frequency and are chosen for reasons of convenience for the Hamiltonian 
of a quantum system under study,
$|\vec X,\vec\theta\,\rangle$ \cite{Korarticle5} is an eigenfunction of the operator
$\vec{\hat X}(\vec\theta)$ with components
$\hat X_\sigma=\hat q_\sigma\cos\theta_\sigma+(\hat p_\sigma\sin\theta_\sigma)/(m_\sigma\omega_\sigma)$
corresponding to the eigenvalue $\vec X$.

For symplectic tomography one can be written 
\be		\label{dequantizerSYMP}
\hat U_M(\vec X,\vec\mu,\vec\nu)=|\vec X,\vec\mu,\vec\nu\,\rangle\langle \vec X,\vec\mu,\vec\nu\,|=
\prod_{\sigma=1}^N
\delta(X_\sigma-\hat q_\sigma\mu_\sigma-\hat p_\sigma\nu_\sigma),
\ee
\be		\label{quantizerSYMP}
\hat D_M(\vec X,\vec\mu,\vec\nu)=
\prod_{\sigma=1}^N\frac{m_\sigma\omega_{\sigma}}{2\pi}
\exp\left\{i\sqrt{\frac{m_\sigma\omega_{\sigma}}{\hbar}}
\left(X_\sigma-\hat q_\sigma\mu_\sigma-\hat p_\sigma\nu_\sigma\right)\right\},
\ee
where $|\vec X,\vec\mu,\vec\nu\,\rangle$ is an eigenfunction of the operator
$\vec{\hat X}(\vec\mu,\vec\nu)$ with components
$\hat X_\sigma=\mu_\sigma\hat q_\sigma+\nu_\sigma\hat p_\sigma$
corresponding to the eigenvalue $\vec X$.

For Wigner representation we have
\be				\label{DeqW}
\hat U_W({\mathbf q},{\mathbf p})=
\frac{1}{(2\pi\hbar)^N}\int \left|{\bf q}-{\bf u}/2\rangle
\exp(-i{\bf p}{\bf u}/\hbar)\langle{\bf q}+{\bf u}/2\right|
{\rm d}^Nu,
\ee
\be			\label{Wigquantizer}
\hat D_W({\mathbf q},{\mathbf p})=2^N
\int{\mathrm d}^Nu\,\exp(2i{\mathbf p}{\mathbf u}/\hbar)|{\mathbf q}+{\mathbf u}\rangle
\langle {\mathbf q}-{\mathbf u}|~;
\ee
for Husimi representation (see \cite{Mizrahi,DavidovichLalovich})
\be				\label{DeqQ}
\hat U_Q({\mathbf q},{\mathbf p})=(2\pi\hbar)^{-N}|\vec\alpha\rangle\langle\vec\alpha|,
~~~~\vec\alpha=\frac{1}{\sqrt2}\left(\sqrt{\frac{m\omega}{\hbar}}{\bf q}+
\frac{i}{\sqrt{\hbar m\omega}}{\bf p}\right),
\ee
where $|{\mathbf q}\rangle$ is an eigenvalue of the position operator,
$|\vec\alpha\rangle$ is a standart boson coherent state,
\bea				
\hat D_Q({\mathbf q},{\mathbf p})&=&\left(\frac{m\omega}{\pi\hbar}\right)^{N/2}
\int{\mathrm d}^Nx{\mathrm d}^Ny\Bigg\{
|{\mathbf x}\rangle\langle {\mathbf y}|
\exp\left(\frac{m\omega}{2\hbar}({\mathbf x}-{\mathbf y})^2\right)
\nonumber \\[3mm]
&\times &
\exp\left[-\frac{m\omega}{\hbar}\left({\mathbf q}-\frac{{\mathbf x}
+{\mathbf y}}{2}\right)^2
-\frac{m\omega}{\hbar} ({\mathbf x} - {\mathbf y})^2
+\frac{i}{\hbar}{\mathbf p}({\mathbf x} - {\mathbf y})
\right]
\nonumber \\[3mm]
&\times &
\prod_{\sigma=1}^{N}\left[
\sum_{n_\sigma=0}^{\infty}\frac{(-1)^{n_\sigma}}{n_\sigma!2^{n_\sigma}}
H_{2n_\sigma}\left(\sqrt{\frac{m\omega}{\hbar}}q_\sigma-
\frac{m\omega}{2\hbar}(x_\sigma+y_\sigma)^2\right)
\right]
\Bigg\}.
\label{Husimiquantizer}
\eea
Likewise, the Glauber-Sudarshan P-function \cite{Glauber,Sudarshan}
(see, also \cite{Mehta1967}) can be introduced with the help of corresponding dequantizer and quantizer
\be		\label{GlaubQD}
\hat U_P(\vec\alpha)=
\left(\frac{e^{|\vec\alpha|^2}}{\pi^{2N}}
\int|\vec\beta\rangle\langle-\vec\beta|e^{|\vec\beta|^2-\vec\beta\vec\alpha^*
+\vec\beta^*\vec\alpha}\mathrm{d}^{2N}\beta\right),
~~~~
\hat D_P(\vec\alpha)=|\vec\alpha\rangle\langle\vec\alpha|~,
\ee
and so on for the other tomographic schemes.

\section{Probability description of spin-1 particles}
\label{Sec3}
In general case the evolution of charged spin-1 particle in the external electro-magnetic
field is determined by Proca equation \cite{Proca1,Proca2}.
This is a relativistic wave  equation of four-component wave function
$(\varphi_0,\,\varphi_1,\,\varphi_2,\,\varphi_3)$.
But in the case of weak relativism it can be reduced to the 
Schr\"odinger type equation for three-component spinor wave function
with the Hermitian Hamiltonian $(\psi_1,\psi_2,\psi_3)$  \cite{Proca3,Proca33}.

For quantum system of charged spin-1 particles without electrical quadrupole moment,
with charge $e$ and mass $m$ in the electro-magnetic field with vector and scalar potentials
${\bf A}({\bf q},t)$, $\varphi({\bf q},t)$ this Hamiltonian has the form:
\be			\label{Hamiltonian1} 
\hat H=\frac{1}{2m}\left(\hat{\bf p}-\frac{e}{c}{\bf A}\right)^2
+e\varphi-\frac{\varkappa}{s}\hat{\bf s}\,{\bf H}=\hat H_0-\frac{\varkappa}{s}\hat{\bf s}\,{\bf H},
\ee
where $\hat H_0$ is an independent on spin part of Hamiltonian, 
${\bf H}=\mbox{rot}{\bf A}$ is a magnetic field, and $\varkappa$ is a magnetic moment of the particle.

The wave function satisfy by the normalization condition
\bdm
\int(|\psi_1(\mathbf{q})|^2+|\psi_2(\mathbf{q})|^2+|\psi_3(\mathbf{q})|^2)
\mathrm{d}^3q=1.
\edm
So, mixed states  are described by the density matrix $\hat\rho_{ij}$
with dimension $3\times3$, which is actually defined by
nine real scalar components.

Analogously with the case of spin-1/2 particles \cite{KorennoyPauli},
to construct the vector portrait of such density matrix 
we must solve the state reconstruction problem, i.e., we have to
find the inverse map, which transforms the set of expectation values  
of observables constituting a quorum to the density matrix.

For this purpose we should choose nine spin-1 states $|\sigma_i,{\mathbf n}_i\rangle$ 
with definite spin projections $\sigma_i$ along the directions ${\mathbf n}_i$, 
which define nine-component dequantizer vector $\vec{\hat{\mathcal{U}}}$ of $3\times3$
spin matrices with components 
$\hat{\mathcal{U}}_i=|\sigma_i,{\mathbf n}_i\rangle\langle\sigma_i,{\mathbf n}_i|$
and quantizer $3\times3$ matrix $\hat{\vec{\mathcal{D}}}$
of nine-component vectors so, that
\be			\label{ortcondition}
\mathrm{Tr}_{kl}\left\{\mathcal{U}_{j(kl)}\mathcal{D}_{(kl)j'}\right\}=
\sum_{k,l=1}^{2s+1}\mathcal{U}_{j(kl)}\mathcal{D}_{(kl)j'}=\delta_{jj'},
~~~
\sum_{j=1}^{(2s+1)^2}\mathcal{U}_{j(kl)}\mathcal{D}_{(k'l')j}=
\delta_{kk'}\delta_{ll'}.
\ee
Here the first index $j=1,2,...,9$ is the number of the component of the nine-component vector,
and $(kl)$ are the indexes of $3\times3$ matrices.
It is obvious that the set of matrixes $\{\hat{\mathcal U}_i\}$ must be linearly independent.
With the help of $\vec{\hat{\mathcal{U}}}$ the nine-component vector
tomogram or quasidistribution is defined as
\be			\label{deffn21}
\vec w(x,\eta,t)=\mbox{Tr}\left\{
\hat\rho(t)\hat U(x,\eta)\otimes\vec{\hat{\mathcal{U}}}
\right\},
\ee
where the trace is calculated also over spin indexes, and $\hat U(x,\eta)$ is defined by formula
(\ref{dequantizerOPT}), (\ref{dequantizerSYMP}), (\ref{DeqW}), (\ref{DeqQ}), or  (\ref{GlaubQD}).
Here $\vec w(x,\eta,t)$ is the aggregate designation of the vector optical
$\vec w(\vec X,\vec\theta,t)$ or symplectic $\vec M(\vec X,\vec\mu,\vec\nu,t)$
tomogram, vector Wigner function $\vec W(\mathbf{q},\mathbf{p},t)$,
vector Husimi function $\vec Q(\mathbf{q},\mathbf{p},t)$, or vector
Glauber-Sudarshan function $\vec P(\vec\alpha,t)$.
For optical and symplectic vector tomograms and for the Husimi vector 
quasidistribution each of the function $w_j(x,\eta,t)$
is the probability distribution of the operator
$\hat{x}(\eta)$ at time $t$ under the condition that the particle has the  corresponding value of 
spin projection along the appropriate direction.
Consequently,   the components of the vector $\vec w(x,\eta,t)$
must be integrable over $\mathrm{d}x$ and
must satisfy the inequalities
\bdm
0\leq w_j(x,\eta,t) \leq 1,~~~~
0\leq \int w_j(x,\eta,t)\mbox{d}x \leq 1,~~~~j=1,...,~9.
\edm
The components of the constructed vector Wigner function and vector Glauber-Sudarshan function
corresponding definite spin projection along the appropriate  direction
are not obligatory non-negative, but definition (\ref{deffn21}) guarantee that they are definitely real.

Such a definition (\ref{deffn21}) for our vector Wigner and Husimi functions
differs from those usually given in literature by many authors,
when the Wigner function $W_{jk}({\bf q},{\bf p},t)$ and
Husimi function $Q_{jk}({\bf q},{\bf p},t)$ become 
$(2s+1)\times(2s+1)$ matrices dependent on position and momentum,
but their non-diagonal elements over the spin indexes are not surely real.
So, the main advantage of such quasidistributions with respect to density matrix 
disappears.

The inverse map of (\ref{deffn21}) is written by means of spin quantizer
$\hat{\vec{\mathcal{D}}}$ as follows
\be			\label{inversemap2}
\hat\rho_{jk}(t)=\int \hat D(x,\eta)\otimes\vec{\mathcal{D}}_{(jk)}\vec w(x,\eta,t)
\mathrm{d}x\mathrm{d}\eta.
\ee

Generalizing equation (\ref{r22_2_22222}) to the case of spin particles we can
write the evolution equation for the components of the tomogram or vector 
quasidistribution
\bea
\partial_t w_j(x,\eta,t)&=&\frac{2}{\hbar}\sum_{k=1}^{(2s+1)^2}
\int \mathrm{Im}\left[\mathrm{Tr}\left\{
\hat U(x,\eta)\otimes\hat{\mathcal{U}}_j \hat H\,\hat D(x',\eta'\,)\otimes
\hat{\mathcal{D}}_k\right\}\right]  \nonumber \\
&&\times w_k(x',\eta',t){\mbox d}x'{\mbox d}\eta'\,,
~~~~j=\overline{1,\,(2s+1)^2}.
\label{eqvectspin}
\eea

Let us point out that the scheme proposed admits 
a generalization to the case of arbitrary spin $s$.
If we have the evolution equation of the quantum system
with spin $s$ for the nonnegative, hermitian, and normalized
density matrix, then we can introduce $(2s+1)^2$-component
vector of $(2s+1)\times(2s+1)$ matrices   dequantizer 
$\vec{\hat{\mathcal U}}$ and dual $(2s+1)\times(2s+1)$
matrix of $(2s+1)^2$-component vectors $\hat{\vec{\mathcal D}}$,
which are related by conditions (\ref{ortcondition}).
After that we can define $(2s+1)^2$-component
vector tomogram (or quasidistribution) in 
accordance with analog of (\ref{deffn21}),
and with the help of the formula analogous to (\ref{inversemap2}) 
we can write the evolution equation of type (\ref{eqvectspin})
for the $(2s+1)^2$-component vector tomogram or quasidistribution.

\section{Example of vector tomography representation
for spin-1 particle}
\label{Sec4}
To determine the dequantizer $\vec{\hat{\mathcal U}}$ we should choose
nine positive projections of the quantum state, which completely define
the density matrix.
Let us choose such spin projectors as follows:
\bea
\vec{\hat{\mathcal U}}&=&\left(\frac{}{}
|s_z=1\rangle\langle s_z=1|,\,
|s_z=0\rangle\langle s_z=0|,\,
|s_z=-1\rangle\langle s_z=-1|,\right.\nonumber \\[3mm]
&&
|s_x=1\rangle\langle s_x=1|,\,
|s_x=0\rangle\langle s_x=0|,\,
|s_{xy}=1\rangle\langle s_{xy}=1|,\nonumber \\[3mm]
&&
\left.|s_{xy}=0\rangle\langle s_{xy}=0|,\,
|s_{yz}=0\rangle\langle s_{yz}=0|,\,
|s_{xz}=0\rangle\langle s_{xz}=0|
\frac{}{}\right),
\label{progection}
\eea
where  $|s_j=\pm1,\,0\rangle$ is an  eigenfunction of the projection of spin operator
to the direction $j$ corresponding to the eigenvalue $\pm1$ or $0$, and
$|s_{xy}\rangle$, $|s_{yz}\rangle$, $|s_{xz}\rangle$ are 
eigenfunctions of  projections of spin operator
to  directions $\vec e_{xy}=(1/\sqrt2,\,1/\sqrt2,\,0)$, $\vec e_{yz}=(0,\,1/\sqrt2,\,1/\sqrt2)$,
$\vec e_{xz}=(1/\sqrt2,\,0,\,1/\sqrt2)$ respectively.

Choose the spin representation, in which components of spin operator
are defined as follows:
\bdm
\hat s_x=\frac{1}{\sqrt2}\left[
\begin{array}{ccc}
0 & 1 & 0 \\
1 & 0 &1 \\
0 & 1 & 0 
\end{array}
\right],~~~~
\hat s_y=\frac{i}{\sqrt2}\left[
\begin{array}{ccc}
0 & -1 & 0 \\
1 & 0 &-1 \\
0 & 1 & 0 
\end{array}
\right],~~~~
\hat s_z=\left[
\begin{array}{ccc}
1 & 0 & 0 \\
0 & 0 & 0 \\
0 & 0 & -1 
\end{array}
\right].
\edm 
After calculations in matrix notations for dequantizer $\vec{\hat{\mathcal U}}$ we have
\bea				
\vec{\hat{\mathcal U}}=
\left\{\hat{\mathcal U}_{j(kl)}\right\}&=&
\left( 
\left[
\begin{array}{ccc}
1 & 0 & 0 \\
0 & 0 & 0 \\
0 & 0 & 0 
\end{array}
\right],\,\,
\left[
\begin{array}{ccc}
0 & 0 & 0 \\
0 &1 & 0 \\
0 & 0 & 0 
\end{array}
\right],\,\,
\left[
\begin{array}{ccc}
0 & 0 & 0 \\
0 & 0 & 0 \\
0 & 0 & 1 
\end{array}
\right],\,\,
\frac{1}{4}\left[
\begin{array}{ccc}
1 & \sqrt2 & 1 \\
\sqrt2 & 2 & \sqrt2 \\
1 & \sqrt2 & 1 
\end{array}
\right],\,\, \right. \nonumber\\ [3mm]
&&\left.
\frac{1}{2}\left[
\begin{array}{ccc}
1 & 0 & -1 \\
0 & 0 & 0 \\
-1 & 0 & 1 
\end{array}
\right],\,\,
\frac{1}{4}\left[
\begin{array}{ccc}
1 & 1-i & -i \\
i+1 & 2 & 1-i \\
i & 1+i & 1 
\end{array}
\right],\,\,
\frac{1}{2}\left[
\begin{array}{ccc}
1 & 0 & i \\
0 & 0 & 0 \\
-i & 0 & 1 
\end{array}
\right],\,\,   \right. \nonumber\\ [3mm]
&&\left.
\frac{1}{4}\left[
\begin{array}{ccc}
1 & i\sqrt2 & 1 \\
-i\sqrt2 & 2 & -i\sqrt2 \\
1 & i\sqrt2 & 1 
\end{array}
\right],\,\,
\frac{1}{4}\left[
\begin{array}{ccc}
1 & -\sqrt2 & -1 \\
-\sqrt2 & 2 & \sqrt2 \\
-1 & \sqrt2 & 1 
\end{array}
\right]
\right).
\label{spinDequv}
\eea
From duality relation (\ref{ortcondition}) after some calculations we obtain 
spin quantizer $\vec{\hat{\mathcal D}}$, which is a $3\times3$
matrix of nine-component vectors
\be				\label{spinDopt}
\vec{\hat{\mathcal D}}=
\left\{\hat{\mathcal D}_{(jk)l}\right\}=
\left[
\begin{array}{ccc}
\vec{\hat{\mathcal D}}_{(11)} & \vec{\hat{\mathcal D}}_{(12)} & \vec{\hat{\mathcal D}}_{(13)} \\
\vec{\hat{\mathcal D}}_{(21)} & \vec{\hat{\mathcal D}}_{(22)} & \vec{\hat{\mathcal D}}_{(23)} \\
\vec{\hat{\mathcal D}}_{(31)} & \vec{\hat{\mathcal D}}_{(32)} & \vec{\hat{\mathcal D}}_{(33)}
\end{array}
\right],
\ee
where $(jk)$ are the indexes of $3\times3$ matrix and $l=1,2,...,9$ \,\,is the index of the component
of nine-component vector
\bea
\vec{\hat{\mathcal D}}_{(11)}&=&\Big(1,\,0,\,0,\,0,\,0,\,0,\,0,\,0,\,0 \Big), \nonumber\\
\vec{\hat{\mathcal D}}_{(12)}&=&\Big( -\frac{1}{2\sqrt2}+i\frac{1-\sqrt2}{2},\,
i\frac{1-\sqrt2}{2},\, -\frac{1}{2\sqrt2}+i\frac{1-\sqrt2}{2}, \,
\frac{1+i}{\sqrt2},\, \nonumber\\
&&\frac{1+i}{\sqrt2},\, -i ,\,-\frac{i}{2},\, \frac{i}{\sqrt2},\,-\frac{1}{\sqrt2} \, \Big),\nonumber\\
\vec{\hat{\mathcal D}}_{(13)}&=&\Big( \frac{1-i}{2}, \,
0,\, \frac{1-i}{2}, \, 0,\,
-1,\, 0,\, i ,\,0,\, 0 \Big),\nonumber\\
\vec{\hat{\mathcal D}}_{(22)}&=&\Big(0,\,1,\,0,\,0,\,0,\,0,\,0,\,0,\,0 \Big),\nonumber\\
\vec{\hat{\mathcal D}}_{(23)}&=&\Big( -\frac{1}{2\sqrt2}+\frac{i}{2}, \,
-\frac{1}{\sqrt2}+\frac{i}{2},\, -\frac{1}{2\sqrt2}+\frac{i}{2}, \,
\frac{1+i}{\sqrt2},\, 0,\, -i ,\,-\frac{i}{2},\, -\frac{i}{\sqrt2},\,\frac{1}{\sqrt2}  \Big), \nonumber\\
\vec{\hat{\mathcal D}}_{(33)}&=&\Big(0,\,0,\,1,\,0,\,0,\,0,\,0,\,0,\,0 \Big), \nonumber
\eea
\bdm
\vec{\hat{\mathcal D}}_{(21)}=\vec{\hat{\mathcal D}}_{(12)}^*,~~~~
\vec{\hat{\mathcal D}}_{(31)}=\vec{\hat{\mathcal D}}_{(13)}^*,~~~~
\vec{\hat{\mathcal D}}_{(32)}=\vec{\hat{\mathcal D}}_{(23)}^*.
\edm
Obviously that  for such of definition (\ref{progection}) of dequantizer $\vec{\hat{\mathcal U}}$, 
three components of the vector $\vec w(x,\eta,t)$
are normalized by the condition
\be			\label{normalization}
\int w_1(x,\eta,t)\mbox{d}x+
\int w_2(x,\eta,t)\mbox{d}x+
\int w_3(x,\eta,t)\mbox{d}x=1.
\ee

For Hamiltonian (\ref{Hamiltonian1}) the evolution equation (\ref{eqvectspin}) of the optical 
vector tomogram is written as follows (see \cite{Korarticle1}):
\be			\label{maineq}
\partial_t\vec w(\vec X,\vec\theta,t)=
\hat{\mathcal M}_w(\vec X,\vec\theta,t)\,\vec w(\vec X,\vec\theta,t)+
\hat{\mathbf S}_w(\vec X,\vec\theta,t)\,\vec w(\vec X,\vec\theta,t),
\ee
where 
\bdm
\hat{\mathcal M}_w(\vec X,\vec\theta,t)=
\frac{2}{\hbar}\mathrm{Im}\,
\hat H_0\left([\hat{\mathbf q}]_w(\vec X,\vec\theta\,),[\hat{\mathbf p}]_w(\vec X,\vec\theta\,),t\right)
\edm
is an operator depending on position $[\hat{\mathbf q}]_w$ and momentum $[\hat{\mathbf p}]_w$
operators  in the optical tomographic representation \cite{Korarticle5}
\be			\label{qFromXTheta}
[\hat q_\sigma]_w(\vec X,\vec\theta)=\sin\theta_\sigma\frac{\partial}{\partial\theta_\sigma}
\left[\frac{\partial}{\partial X_\sigma}\right]^{-1}
+X_\sigma\cos\theta_\sigma+i\frac{\hbar\sin\theta_\sigma}
{2m_\sigma\omega_{\sigma}}
\frac{\partial}{\partial X_\sigma},
\ee
\be			\label{pFromXTheta}
[\hat p_\sigma]_w(\vec X,\vec\theta)=m\omega_{\sigma}\left(-\cos\theta_\sigma\left[\frac{\partial}{\partial X_\sigma}\right]^{-1}
\frac{\partial}{\partial\theta_\sigma}+X_\sigma\sin\theta_\sigma\right)-\frac{i\hbar}{2}
\cos\theta_\sigma\frac{\partial}{\partial X_\sigma},
\ee
and $\hat{\mathbf S}_w(\vec X,\vec\theta,t)$ is a $9\times9$ matrix operator,
responsible for the interaction of spin with the magnetic field
\be			\label{genSmatrix}
\hat{\mathbf S}_w(\vec X,\vec\theta,t)=-\frac{2\varkappa}{\hbar s}\mathrm{Im}
\left\{
\sum_{l,m,m'=1}^{2s+1}\mathcal{U}_{j(lm)}
\left[\hat{\bf s}\,{\bf H}\left([\hat{\mathbf q}]_w(\vec X,\vec\theta\,)\right)
\right]_{(mm')}\mathcal{D}_{(m'l)k}
\right\}.
\ee
With omitted arguments and introduced designations
\bdm
[\hat A_j]_w=A_j\left([\hat{\mathbf q}]_w(\vec X,\vec\theta\,),t\right),
~~~~
\tilde{\mathrm H}_j=
[\hat {\mathrm H}_j]_w=\mathrm H_j\left([\hat{\mathbf q}]_w(\vec X,\vec\theta\,),t\right),
\edm
\bdm
\left[\nabla_{\mathbf q}\hat{\mathbf A}\right]_w
=\nabla_{\mathbf q}{\mathbf A}\left(
{\mathbf q}\rightarrow [\hat{\bf q}]_w(\vec X,\vec\theta),t\right)
\edm
the explicit form of  $\hat{\mathcal M}_w$ 
in general case of time-dependent and non-homogeneous electromagnetic field
is written as
\bea
\hat{\mathcal M}_w(\vec X,\vec\theta,t)&=&
\sum_{n=1}^{3}\omega_{n}\left[\cos^2\theta_n\frac{\partial}{\partial\theta_n}
-\frac{1}{2}\sin2\theta_n\left\{1+X_n\frac{\partial}{\partial X_n}\right\}\right] 
+\frac{2e}{\hbar}\,\mathrm{Im}\,[\hat{\varphi}]_w
\nonumber \\[3mm]
&+&\frac{e^2}{mc^2\hbar}\,\mathrm{Im}[\hat{\bf A}]_w^2
-\frac{2e}{mc\hbar}\,\mathrm{Im}\left[\hat{\mathbf A} \hat{\mathbf p}\right]_w
+\frac{e}{mc}\,\mathrm{Re}\left[{\nabla_{\mathbf q}{\mathbf A}}\right]_w.
\label{explicitM}
\eea

For symplectic   vector tomography 
we can find the evolution equation
\be                             \label{dynEqSympSpin}
\partial_t\vec M({\vec X},{\vec\mu},\vec\nu,t)=
\hat{\mathcal M}_M({\vec X},{\vec\mu},\vec\nu,t)\,\vec M({\vec X},{\vec\mu},\vec\nu,t)
+\hat{\mathbf S}_M(\vec X,\vec\mu,\vec\nu,t)\,\vec M(\vec X,\vec\mu,\vec\nu,t),
\ee
where operator $\hat{\mathcal M}_M({\vec X},{\vec\mu},\vec\nu,t)$ corresponds to spinless part $\hat H_0$
of the Hamiltonian (\ref{Hamiltonian1})
\bea
\hat{\mathcal M}_M({\vec X},{\vec\mu},\vec\nu,t)&=&\frac{2}{\hbar}\,
\mathrm{Im}\,\hat H_0\left([\hat{\mathbf p}]_M({\vec X},{\vec\mu},\vec\nu),
[\hat{\mathbf q}]_M({\vec X}{\vec\mu},\vec\nu),t\right)
=\vec\mu\frac{\partial}{\partial\vec\nu}
+\frac{2e}{\hbar}\,\mathrm{Im}\,[\hat{\varphi}]_M
\nonumber \\[3mm]
&+&\frac{e^2}{mc^2\hbar}\,\mathrm{Im}[\hat{\mathbf A}]_M^2
-\frac{2e}{mc\hbar}\,\mathrm{Im}\left[\hat{\mathbf A} \hat{\mathbf p}\right]_M
+\frac{e}{mc}\,\mathrm{Re}\left[\nabla_{\mathbf q}{\mathbf A}\right]_M,
\eea
where
\bdm
[\hat A_j]_M=A_j\left([\hat{\mathbf q}]_M(\vec X,\vec\mu,\vec\nu\,),t\right),~~~~
[\hat\varphi]_M=\varphi\left([\hat{\mathbf q}]_M(\vec X,\vec\mu,\vec\nu\,),t\right),
\edm
\bdm
[\nabla_{\mathbf q}{\mathbf A}]_M=\nabla_{\mathbf q}{\mathbf A}\left(
{\mathbf q}\rightarrow [\hat{\mathbf q}]_M(\vec X,\vec\mu,\vec\nu\,),t\right), 
\edm
and $[\hat{\mathbf q}]_M$, $[\hat{\mathbf p}]_M$ are position
and momentum operators (\ref{defqpSymp}) in the symplectic representation (see \cite{Korarticle2})
\bea
[\hat p_\sigma]_M&=&\left(-\left[\frac{\partial}{\partial X_\sigma}\right]^{-1}
\frac{\partial}{\partial\nu_\sigma}-i\frac{\mu_\sigma\hbar}{2}\frac{\partial}{\partial X_\sigma}
\right),\nonumber \\[3mm]
[\hat q_\sigma]_M&=&\left(-\left[\frac{\partial}{\partial X_\sigma}\right]^{-1}
\frac{\partial}{\partial\mu_\sigma}+i\frac{\nu_\sigma\hbar}{2}\frac{\partial}{\partial X_\sigma}
\right).
\label{defqpSymp}
\eea
The $9\times9$ matrix operator $\hat{\mathbf S}_M(\vec X,\vec\mu,\vec\nu,t)$ is defined by
the similar  formula (\ref{genSmatrix}), where the operators of components
of the magnetic field $\tilde{\rm H}_j$ must be replaced with corresponding
operators in the symplectic tomography representation
$[\hat{\mathrm H}_j]_M={\mathrm H}_j\left([\hat{\mathbf q}]_M(\vec X,\vec\mu,\vec\nu\,),t\right)$.

Making similar calculation we can obtain such evolution equation 
for our  vector Wigner function, which is  a generalization of the
Moyal equation \cite{Moyal1949}
\bea			
\frac{\partial }{\partial t}\vec W({\mathbf q},{\mathbf p},t)&=&
\left[-\frac{\mathbf p}{m}
\frac{\partial}{\partial{\mathbf q}}
+\frac{2e}{\hbar}\,\mbox{Im}\,{\varphi}
\left(
{\bf q}+\frac{i\hbar}{2} \frac{\partial}{\partial {\bf p}},t
\right)
+\frac{e^2}{mc^2\hbar}\,\mathrm{Im}{\bf A}^2
\left(
{\bf q}+\frac{i\hbar}{2} \frac{\partial}{\partial {\bf p}},t
\right)\right.
 \nonumber \\[3mm]
&+&\left.
-\frac{2e}{mc\hbar}\,{\mathrm{Im}}\left\{{\mathbf A}
\left(
{\mathbf q}+\frac{i\hbar}{2} \frac{\partial}{\partial {\mathbf p}},t
\right)
\left(
{\mathbf p}-\frac{i\hbar}{2} \frac{\partial}{\partial {\mathbf q}}
\right)
\right\}\right.  \nonumber \\[3mm]
&+&\left.\frac{e}{mc}\,\mathrm{Re}{\nabla_{\bf q}{\bf A}}
\left(
\mathbf q \to {\mathbf q}+\frac{i\hbar}{2} \frac{\partial}{\partial {\mathbf p}},t
\right)
+\hat{\mathbf S}_W({\bf q},{\bf p},t)
 \right] \vec W({\mathbf q},{\mathbf p},t),
\label{Moyal2}
\eea
where $9\times9$ matrix operator $\hat{\mathbf S}_W({\bf q},{\bf p},t)$
is defined by the same formula 
(\ref{genSmatrix}), where the operators of components
of the magnetic field $\tilde{\rm H}_j$ must be replaced with corresponding
operators in the Wigner representation
${\mathrm H}_j\left( {\mathbf q}+\frac{i\hbar}{2} \frac{\partial}{\partial {\mathbf p}},t \right)$.

The corresponding generalization of the evolution equation of the Husimi function 
\cite{Mizrahi2} to the case of vector quasidistribution has the form 
(for simplicity we choose the system of measurements so that $m=\omega=\hbar=1$):
\bea			
\frac{\partial }{\partial t}\vec Q({\mathbf q},{\mathbf p},t)&=&
\left[-{\mathbf p}
\frac{\partial}{\partial{\mathbf q}}-\frac{1}{2}\frac{\partial}{\partial{\mathbf q}}
\frac{\partial}{\partial{\mathbf p}}
+\frac{2e}{\hbar}\,\mbox{Im}\,{\varphi}
\left(
{\bf q}+\frac{1}{2}\frac{\partial}{\partial{\mathbf q}}
+\frac{i}{2} \frac{\partial}{\partial {\bf p}},t
\right)\right.  \nonumber \\[3mm]
&+&\frac{e^2}{c^2}\,\mathrm{Im}{\bf A}^2
\left(
{\bf q}
+\frac{1}{2}\frac{\partial}{\partial{\mathbf q}}
+\frac{i}{2} \frac{\partial}{\partial {\bf p}},t
\right)
 \nonumber \\[3mm]
&-&\frac{2e}{c}\,{\mathrm{Im}}\left\{{\mathbf A}
\left(
{\mathbf q}
+\frac{1}{2}\frac{\partial}{\partial{\mathbf q}}
+\frac{i}{2} \frac{\partial}{\partial {\mathbf p}},t
\right)
\left(
{\mathbf p}+\frac{1}{2}\frac{\partial}{\partial{\mathbf p}}
-\frac{i}{2} \frac{\partial}{\partial {\mathbf q}}
\right)
\right\}  \nonumber \\[3mm]
&+&\left.\frac{e}{c}\,\mathrm{Re}{\nabla_{\bf q}{\bf A}}
\left(
\mathbf q \to {\mathbf q}
+\frac{1}{2}\frac{\partial}{\partial{\mathbf q}}
+\frac{i}{2} \frac{\partial}{\partial {\mathbf p}},t
\right)+\hat{\mathbf S}_Q({\bf q},{\bf p},t)
 \right] \vec Q({\mathbf q},{\mathbf p},t), \nonumber \\
\label{MoyalHusimi}
\eea
where $9\times9$ matrix operator $\hat{\mathbf S}_Q({\bf q},{\bf p},t)$
is defined by (\ref{genSmatrix}) in which components
of the magnetic field $\tilde{\rm H}_j$ are  replaced with 
${\mathrm H}_j\left( {\mathbf q}
+\frac{1}{2}\frac{\partial}{\partial{\mathbf q}}
+\frac{i}{2} \frac{\partial}{\partial {\mathbf p}},t \right)$.

\section{Conclusion}
\label{Sec5}
To resume we point out the main results of our paper.
We have constructed the nine-component positive vector optical 
tomographic probability portrait of quantum state of spin-1 particles,
which contains full spatial and spin information about state
without redundancy. We have expanded suggested approach to symplectic tomography representation and 
to representations with quasidistributions.
All of the components of the constructed vector Wigner function are real,
and all of the components of the vector Husimi function are non-negative.

We found the evolution equations for such vector optical
and symplectic tomograms and vector quasidistributions for arbitrary Hamiltonian
and obtained these equations in special case of the quantum system of 
charged spin-1 particle in arbitrary electro-magnetic field,
which are analogs of non-relativistic Proca equation in appropriate representations.
Also we discussed the generalization of our approach to the cases of arbitrary spin.

The general equations obtained  are relatively complicated, but in many 
special cases they are much simpler and could allow for the possibility 
of analytical and numerical solutions.

The results of the paper explicitly demonstrate the possibility of formulation
of quantum mechanics of the systems with spins in terms of joint probability distributions
without application of wave functions or density matrixes.



\begin{thebibliography}{99}
\bibitem{BerBer} Bertrand, J., Bertrand, P.:  A tomographic approach to Wigner's function.
 Found.~Phys. \textbf{17}, 397 (1987)
\bibitem{VogRis} Vogel, K., Risken,   H.: Determination of quasiprobability distributions
 in terms of probability distributions for the rotated quadrature phase. Phys. Rev. A \textbf{40}, 2847(R) (1989)
\bibitem{Mancini96} Mancini, S., Man'ko, V.I., Tombesi,  P.: Symplectic tomography as classical approach
to quantum systems. Phys.~Lett.~A \textbf{213}, 1 (1996)
\bibitem{ManciniFoundPhys97} Mancini, S., Man'ko, V. I., Tombesi, P.: Classical-like description of quantum 
 dynamics by means of symplectic tomography. Found.~Phys. \textbf{27}, 801 (1997)
\bibitem{IbortPhysScr} Ibort, A., Man'ko,  V.I.,  Marmo, G., Simoni,  A., Ventriglia, F.: 
 An introduction to the tomographic picture of quantum mechanics. Phys.~Scr. \textbf{79}, 065013 (2009)
\bibitem{Korarticle2} Korennoy, Ya.A., Man'ko, V.I.: Probability representation of the quantum 
 evolution and energy-level equations for optical tomograms.
 J.~Russ.~Laser~Res. \textbf{32}, 74 (2011)
\bibitem{Korarticle1} Korennoy, Ya.A., Man'ko, V.I.: Evolution equation of the optical tomogram for 
 arbitrary quantum Hamiltonian and optical tomography of relativistic classical
 and quantum systems.  J.~Russ.~Laser~Res. \textbf{32}, 338 (2011)
\bibitem{NewtonYoung} Newton, R.G., Young, B.: Measurability  of the spin density  matrix.
 Ann.~Phys. \textbf{49}, 393 (1968)
\bibitem{DodonovManko1997} Dodonov, V.V., Man'ko, V.I.: Positive distribution description for spin states.
 Phys.~Lett.~A \textbf{229}, 335 (1997)
\bibitem{Weigert2000} Weigert, S.: Quantum time evolution in terms of nonredundant probabilities.
 Phys.~Rev.~Lett. \textbf{84}, 802 (2000)
\bibitem{Weigert1992} Weigert, S.: Pauli problem for a spin of arbitrary length:
 A simple method to determine its wave function. Phys.~Rev.~A \textbf{45}, 7688 (1992)
\bibitem{AmietWeigert9} Amiet, J.-P., Weigert, S.: Reconstructing the density matrix of a spin $s$
 through Stern - Gerlach measurements. J.~Phys.~A: Math.~Gen. \textbf{31}, L543 (1998)
\bibitem{AmietWeigert10} Amiet, J.-P., Weigert, S.: Reconstructing a pure state of a spin $s$
 through three Stern-Gerlach measurements. J.~Phys.~A: Math.~Gen. \textbf{32}, 2777 (1999)
\bibitem{AmietWeigert11} Amiet, J.-P., Weigert, S.: Coherent states and the reconstruction of pure spin states.
 J.~Opt.~B: Quantum~Semiclass.~Opt. \textbf{1}, L5 (1999)
\bibitem{AmietWeigert12} Amiet, J.-P., Weigert, S.: Reconstructing the density matrix of a spin $s$
 through Stern-Gerlach measurements: II. J.~Phys.~A: Math.~Gen. \textbf{32}, L269 (1999)
\bibitem{AmietWeigert14} Amiet, J.-P., Weigert, S.: Discrete $Q$- and $P$-symbols for spin $s$.
 J.~Opt.~B: Quantum~Semiclass.~Opt. \textbf{2}, 118 (2000)
\bibitem{HeissWeigert2000} Heiss, S., Weigert, S.: Discrete Moyal-type representations for a spin.
 Phys.~Rev.~A \textbf{63}, 012105 (2000)
\bibitem{FilippovManko2010} Filippov, S.N., Man'ko, V.I.:  Inverse spin-$s$ portrait and representation
of qudit states by single probability vectors.
 J.~Russ.~Laser~Res. \textbf{31}, 32 (2010)
\bibitem{MankoMarmo2004} Man'ko, V.I., Marmo, G., Sudarshan, E. C. G., Zaccaria, F.: 
 Positive maps of density matrix and a tomographic criterion of entanglement. 
 Phys.~Lett.~A \textbf{327}, 353 (2004)
\bibitem{ManciniOlgaManko2001} Mancini, S., Man'ko, O.V., Man'ko, V.I., Tombesi, P.:
 The Pauli equation for probability distributions. J.~Phys.~A: Math.~Gen. \textbf{34}, 3461 (2001)
\bibitem{KorennoyPauli} Korennoy, Ya.A., Man'ko, V.I.: Pauli equation 
 for joint tomographic probability distribution. arXiv: 1412.7873 (2014)
\bibitem{Korarticle5} Amosov, G. G., Korennoy, Ya.A., Man'ko, V.I.: Description and measurement
 of observables in the optical tomographic probability representation of quantum mechanics.
 Phys.~Rev.~A \textbf{85}, 052119 (2012)
\bibitem{Wigner32} Wigner, E.: On the quantum correction for thermodynamic equilibrium.
 Phys.~Rev. \textbf{40}, 749, (1932)
\bibitem{Husimi40} Husimi, K.: Some  formal properties of the density matrix.
 Proc.~Phys.-Math.~Soc.~Japan \textbf{22}, 264 (1940)
\bibitem{Glauber} Glauber, R.J.: Photon correlations.
 Phys.~Rev.~Lett. \textbf{10}, 84 (1963)
\bibitem{Sudarshan} Sudarshan, E.C.G.: Equivalence of semiclassical and quantum mechanical
 descriptions of statistical light beams. Phys.~Rev.~Lett. \textbf{10}, 277 (1963)
\bibitem{SIGMA10(2014)086} Lizzi, F., Vitale, P.: Matrix bases for star products: a review.
 SIGMA \textbf{10}, 086 (2014) 
\bibitem{Moyal1949} Moyal, J.E.: Quantum mechanics as a statistical theory.
 Proc.~Cambrige~Philos.~Soc. \textbf{45}, 99 (1949)
\bibitem{Mizrahi} Mizrahi, S.S.: Quantum mechanics in the Gaussian wave-packet
 phase space representation. Physica~A \textbf{127}, 241 (1984)
\bibitem{DavidovichLalovich} Davidovi\'c, D.M., Lalovi\'c, D.: When does a given function  
 in phase space belong to the class of Husimi  distributions? 
 J.~Phys.~A: Math.~Gen. \textbf{26}, 5099 (1993)
\bibitem{Mehta1967} Mehta, C.L.: Diagonal coherent-state representation of quantum operators.
 Phys.~Rev.~Lett. \textbf{18}, 752 (1967)
\bibitem{Proca1} Proca, A.: Sur la th\'eorie ondulatoire des \'electrons positifs et n\'egatifs.
 J.~Phys.~Radium \textbf{7}, 347 (1936)
\bibitem{Proca2} Proca, A.:  Sur les equations fondamentales des particules el\'ementaires.
 C.~R.~Acad.~Sci.~Paris \textbf{202}, 1366 (1936)
\bibitem {Proca3} Proca, A.: Th\'eorie non relativiste des particules \'a spin entier.
 J.~Phys.~Radium \textbf{9}, 61 (1938)
\bibitem{Proca33} Obukhov, I.A., Peres-Fernandes, V.K.,  Khalilov, V.R.:
 Vector field equations in external electro-magnetic fields. 
 Russ.~Phys.~Journal \textbf{26}, 1117 (1983)
\bibitem{Mizrahi2} Mizrahi, S.S.: Quantum mechanics in the Gaussian wave-packet
 phase space representation II: dynamics. Physica~A \textbf{135}, 237 (1986)



\end{thebibliography}
\end{document}